\documentclass{article}
\usepackage{spconf}

\usepackage[utf8]{inputenc} 
\usepackage[T1]{fontenc}    
\usepackage{hyperref}       
\usepackage{url}            
\usepackage{booktabs}       
\usepackage{amsfonts}       
\usepackage{nicefrac}       
\usepackage{microtype}      
\usepackage{amsmath}
\usepackage{amssymb,bm,graphicx}
\usepackage[ruled, vlined, linesnumbered]{algorithm2e}
\usepackage[color=green]{todonotes}

\DeclareMathOperator{\vectorize}{vec}

\DeclareMathOperator{\tr}{tr}

\DeclareMathAlphabet\mathcalbf{OMS}{cmsy}{b}{n}
\usepackage{xcolor}         

\usepackage[font=small]{caption}
\usepackage{amsmath}
\usepackage{amssymb}
\usepackage{amsthm}

\usepackage{subcaption} 
\usepackage{booktabs}
\usepackage[ruled,vlined]{algorithm2e}
\usepackage{algpseudocode}
\usepackage{scalerel}
\usepackage{verbatim}
\usepackage{commath}
\SetCommentSty{mycommfont}

\algdef{SE}{Begin}{End}{\textbf{begin}}{\textbf{end}}

\theoremstyle{definition}
\newtheorem{remark}{Remark}
\newtheorem{example}{Example}

\title{Graph-Regularized Tensor Regression: A Domain-Aware Framework \\ for Interpretable  Multi-Way Financial Modelling}

\name{Yao Lei Xu, Kriton Konstantinidis, Danilo P. Mandic}
\address{
    Dept. of Electrical and Electronic Engineering, Imperial College London, London SW7 2AZ, UK\\
    E-mails: \{yao.xu15, k.konstantinidis19, d.mandic\}@imperial.ac.uk
}

\begin{document}

\maketitle

\begin{abstract}
Analytics of financial data is inherently a Big Data paradigm, as such data are collected over many assets, asset classes, countries, and time periods. This represents a challenge for modern machine learning models, as the number of model parameters needed to process such data grows exponentially with the data dimensions; an effect known as the Curse-of-Dimensionality. Recently, Tensor Decomposition (TD) techniques have shown promising results in reducing the computational costs associated with large-dimensional financial models while achieving comparable performance. However, tensor models are often unable to incorporate the underlying economic domain knowledge. To this end, we develop a novel Graph-Regularized Tensor Regression (GRTR) framework, whereby knowledge about cross-asset relations is incorporated into the model in the form of a graph Laplacian matrix. This is then used as a regularization tool to promote an economically meaningful structure within the model parameters. By virtue of tensor algebra, the proposed framework is shown to be fully interpretable, both coefficient-wise and dimension-wise. The GRTR model is validated in a multi-way financial forecasting setting and compared against competing models, and is shown to achieve improved performance at reduced computational costs. Detailed visualizations are provided to help the reader gain an intuitive understanding of the employed tensor operations. 


\end{abstract}


\section{Introduction} \label{sec:intro}

Financial data are naturally multi-dimensional and multi-variate; indeed, a typical financial data setting involves various market data features which span over time and across several assets. It therefore occurs routinely that when considering long time horizons, many assets, various geographies, and asset classes, the dimensionality of each mode can quickly explode. This results in a large number of model parameters which renders the training prone to over-fitting and impedes efficient learning, a phenomenon known as the \textit{Curse of Dimensionality} \cite{Cichocki2014}.

An emerging framework which is suited to deal with the Curse of Dimensionality associated with financial data are tensor-based models \cite{201712, xu2021ttrnn, xu2020multigraph}. Tensors are multi-dimensional generalization of vectors (order-1 tensors) and matrices (order-2 tensors). In particular, Tensor Decomposition (TD) techniques, similar to their matrix decomposition counterparts, have the ability to logarithmically compress large multi-dimensional tensors into an efficient low-rank format by exploiting intra- and inter-dependencies across data dimensions (tensor modes) \cite{cichocki2016tensor, Mandic2017, TDforSP}. This both preserves the underlying data structure and interpretability, while bypassing the Curse-of-Dimensionality, as the associated computational costs reduce from an original exponential one to a linear one in the data dimensions. 

In addition, financial data exhibit irregular relational structures, which cannot be handled via traditional machine learning techniques. For instance, pairs of stocks can exhibit varying degrees of correlations, depending on their industries, market capitalizations, and value factors, but to mention a few. Currencies also exhibit varying relational structures depending on the interest rate differential and other macroeconomic factors between pairs of countries. Such irregular relational structures, which are well understood and researched within the quantitative finance community, cannot be straightforwardly integrated into standard machine learning methods, resulting in sub-optimal models. 

To this end, Graph Signal Processing techniques have recently gained the interest of the data analytics community, as they generalize traditional machine learning techniques to data on irregular domains \cite{stankovic2019graph, stankovic2019graphII, stankovic2020graphIII, stankovic2019understanding}, achieving state-of-the-art performance in tasks where the underlying data relations can be captured as a graph, including financial modelling \cite{xu2020multigraph, wang2021review}. In this way, assets can be represented as nodes of a graph, while the edges connecting different nodes may be weighted depending on their relational information. 

Financial modelling would therefore greatly benefit from a joint consideration of tensor and graph techniques, to simultaneously overcome the Curse of Dimensionality inherent to financial data while benefiting from domain knowledge about the underlying relational structures. To this end, we introduce a novel Graph-Regularized Tensor Regression (GRTR) framework, which allows for the regression of large multi-dimensional data while efficiently incorporating prior domain knowledge into the model parameter estimation. In addition to improved performance and reduced space complexity costs, we show that such a model is fully interpretable, both in the coefficient-wise sense and in the mode-wise sense. Particular effort has been made to provide intuitive understanding of the employed techniques through detailed visualizations of the underlying operations using Tensor Network (TN) diagrams, to provide readers with an insight into the virtues of tensor algebra when dealing with multiple data dimensions.

The rest of the paper is organized as follows. Related work is first discussed in Section \ref{sec:related_work}. The tensor and graph preliminaries necessary to follow this work are presented in Section \ref{sec:prelim}. Next, we introduce the proposed Graph Regularised Tensor Regression (GRTR) framework, derive the corresponding learning algorithm, and analyse the associated space and time complexities in Section \ref{sec:grtr}. The proposed model is validated and compared against traditional algorithms in a synthetic data based experiment in Section \ref{sec:exp_syn}, as well as in a real data based financial forecasting experiment in Section \ref{sec:exp}. The interpretability of the trained model is then discussed in detail in Section \ref{sec:interpretability}. Finally, we conclude and summarize our findings in Section \ref{sec:conc}. 

\section{Related Work} \label{sec:related_work}

The proposed framework can be considered  under the tensor regression paradigm \cite{zhou2013tensor,MAL-087} whereby, contrary to standard tensor regression, graph regularization  makes it possible to incorporate domain knowledge within the financial modelling setting. This is different from current graph regularized tensor decomposition models \cite{sofuoglu2019graph,article12, 9058984}, which have been used to only help preserve local relationships among tensor samples. In other words, whereas such models are concerned with tensor representations of multi-dimensional data, we are concerned with regression tasks. The recently introduced \textit{Graph Tensor Networks} framework \cite{xu2022attention, xu2020recurrent, xu2020multigraph} also employs graphs within the tensor network paradigm, however, it is a black-box model with graphs serving as an added filter, whereas in this work we employ graph regularization as a mathematical constraint. This allows us to maintain model interpretability throughout the data processing chain, while identifying and promoting meaningful relationships between the regression variables. The proposed GRTR framework is also different from Tensor Decomposition based kernel regression methods discussed in \cite{stoudenmire2016supervised, 9397437}, as hand-crafted feature mappings are not required. Research on the use of tensors in finance includes the work in \cite{xu2021ttrnn,li2015tensor,spelta2017financial,201712}, which elaborate on the benefits of tensors as an ideal data structure to compactly represent the inherently multi-dimensional financial data. Tensor decomposition were therefore a natural choice to form the backbone of the proposed GRTR framework.

\section{Preliminaries} \label{sec:prelim}

This section introduces the necessary background for this work. We refer the readers to \cite{cichocki2016tensor} and \cite{stankovic2019graph} for an in-depth overview of tensor algebra and graph signal processing, 

\subsection{Tensors and Basic Tensor Operations} \label{sec:prelim_tensors}

\noindent\textbf{Definitions.}
    A real-valued tensor is a multidimensional array, denoted by a calligraphic font, e.g., $\mathcalbf{X}\in\mathbb{R}^{I_1\times\dots\times I_N}$, where $N$ is the order of the tensor (i.e., number of dimensions), and $I_n$ ($1 \leq n \leq N$) is the size (i.e., the dimensionality) of its $n$\textsuperscript{th} mode. Matrices (denoted by bold capital letters, e.g., $\mathbf{X}\in\mathbb{R}^{I_1\times I_2}$) can be seen as second order tensors ($N=2$), vectors are denoted by bold lower-case letters, e.g., $\mathbf{x}\in\mathbb{R}^{I}$, and can be seen as order-1 tensors ($N=1$), and scalars (denoted by lower-case letters, e.g., $x\in\mathbb{R}$) are tensors of order $N=0$. A specific entry of a tensor $\mathcalbf{X}\in\mathbb{R}^{I_1\times\dots\times I_N}$ is given by $x_{i_1,\dots,i_N}\in\mathbb{R}$. The following conventions for basic linear/multilinear operations are employed throughout the paper. 

\begin{figure}[t]
	\centering
	\includegraphics[width=1.0\linewidth]{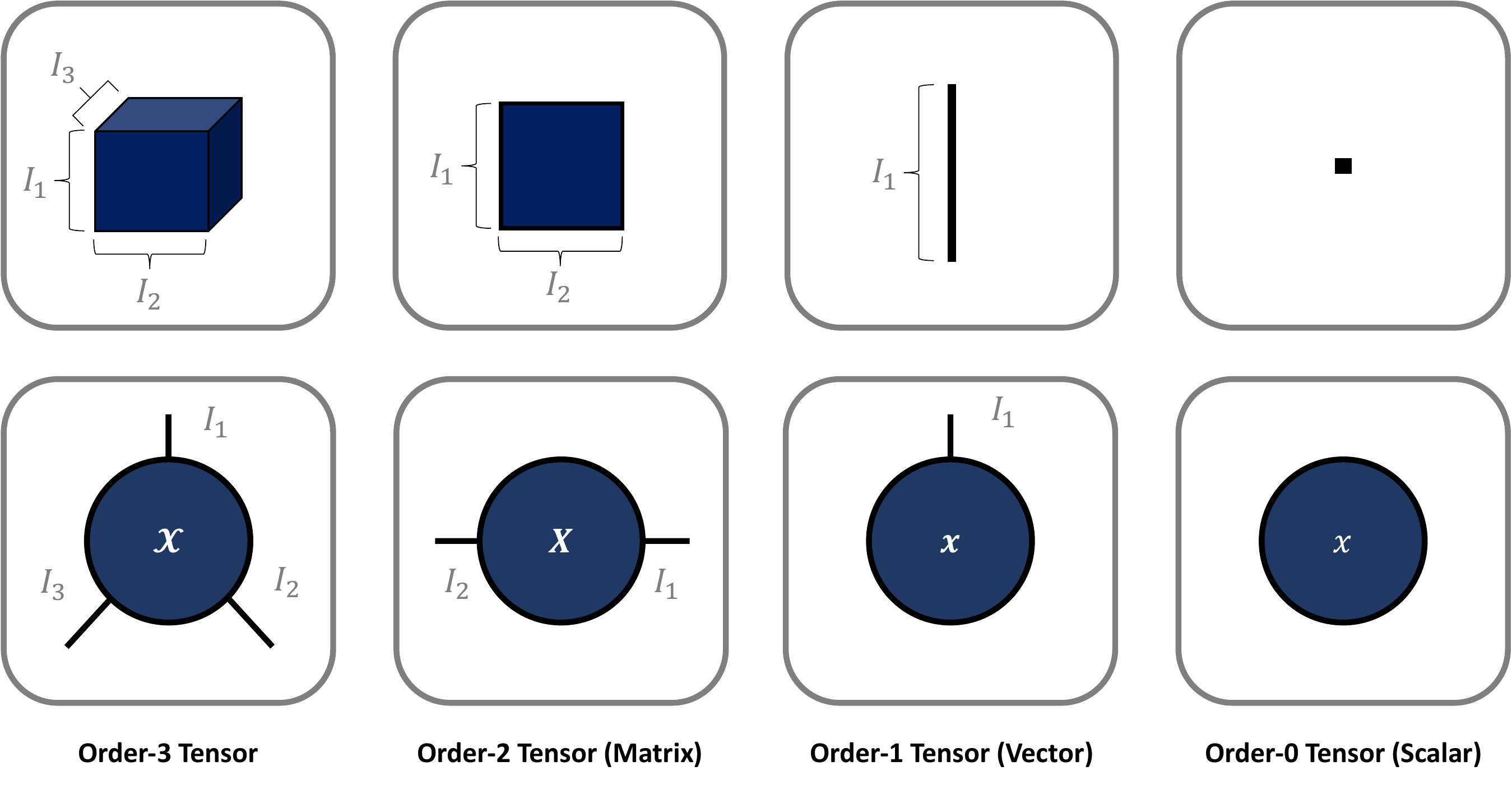}
	\caption{Order-3 tensors, matrices, vectors, and scalars illustrated in the standard form (top) and the tensor network form (bottom). The gray symbols, $I_1$, $I_2$. $I_3$, denote the size of each mode.}
	\label{fig:basic_tensors}
\end{figure}

\begin{figure}[t]
	\centering
	\includegraphics[width=1.0\linewidth]{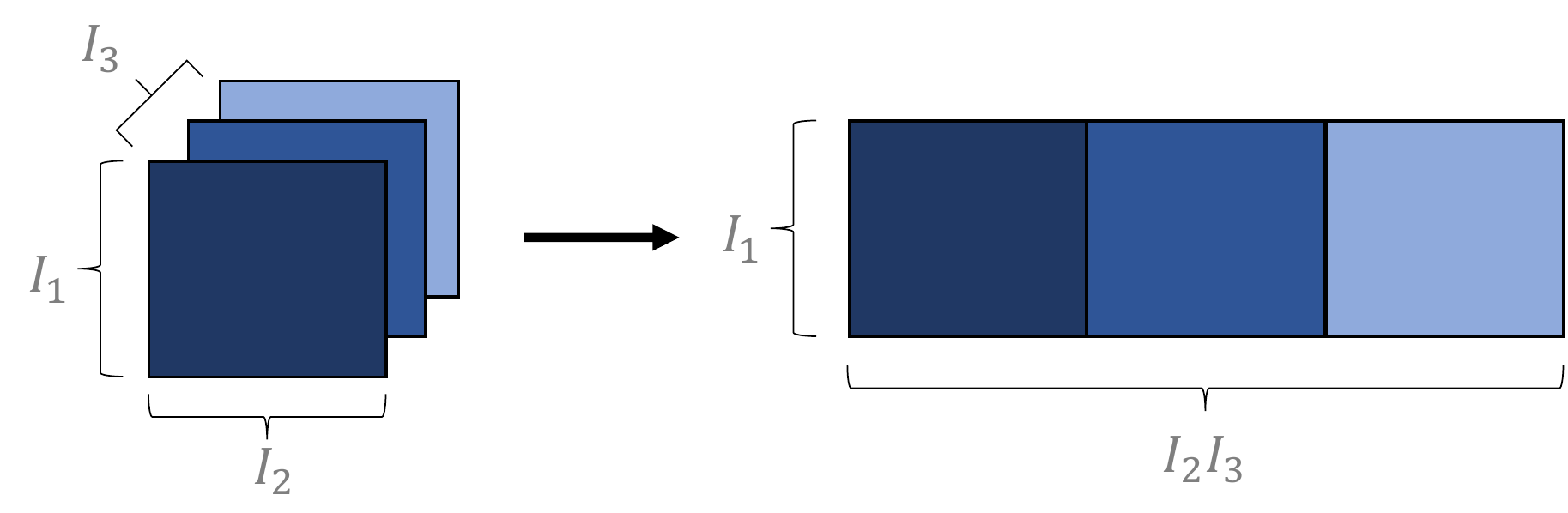}
	\caption{Mode-1 matricization (unfolding) of an order-3 tensor, whereby the front slices (i.e., collection of mode-1 columns vectors) are stacked to form a short and wide matrix.} 
	\label{fig:unfolding}
\end{figure}

\noindent\textbf{Vectorization and matricization.}
    Given an order-$N$ tensor, $\mathcalbf{X}\in\mathbb{R}^{I_1\times\cdots\times I_N}$, its \textit{vectorization} reshapes the multidimensional array into a large vector, $\mathbf{x} \in \mathbb{R}^{I_1 I_2 \cdots I_N}$. The mode-$n$ \textit{matricization} unfolds the tensor into a matrix, $\mathbf{X}_{(n)}\in\mathbb{R}^{I_n\times I_1I_2\cdots I_{n-1}I_{n+1}\cdots I_N}$, as illustrated in Figure \ref{fig:unfolding}.
   
\noindent\textbf{Outer product.}
    The \textit{outer product} of vectors $\mathbf{a}\in\mathbb{R}^{I}$ and $\mathbf{b}\in\mathbb{R}^{J}$ is given by $\mathbf{c}=\mathbf{a}\circ\mathbf{b}\in\mathbb{R}^{I\times J}$, with $c_{i,j}=a_ib_j$.
        
\noindent\textbf{Kronecker product.}
    The \textit{Kronecker product} of two matrices $\mathbf{A}\in\mathbb{R}^{I\times J}$ and $\mathbf{B}\in\mathbb{R}^{K\times L}$ is denoted by $\mathbf{C}=\mathbf{A}\otimes \mathbf{B}\in\mathbb{R}^{IK\times JL}$, with $c_{(i-1)K+k,(j-1)L+l}=a_{i,j}b_{k,l}$.   


\noindent\textbf{Khatri-Rao product.}
    The \textit{Khatri-Rao product} of two matrices $\mathbf{A}=[\mathbf{a}_1,\dots,\mathbf{a}_R]\in\mathbb{R}^{I\times R}$ and $\mathbf{B}=[\mathbf{b}_1,\dots,\mathbf{b}_R]\in\mathbb{R}^{J\times R}$ is denoted by $\mathbf{C}=\mathbf{A}\odot \mathbf{B}\in\mathbb{R}^{IJ\times R}$, where the columns $\mathbf{c}_r=\mathbf{a}_r\otimes \mathbf{b}_r$, $1\leq r\leq R$.

\noindent\textbf{Tensor contraction.}
    The \textit{contraction} of an $N$\textsuperscript{th}-order tensor, $\mathcalbf{A}\in\mathbb{R}^{I_1\times \dots \times I_N}$, and an $M$\textsuperscript{th}-order tensor $\mathcalbf{B}\in\mathbb{R}^{J_1\times \dots \times J_M}$, over the $n$\textsuperscript{th} and $m$\textsuperscript{th} modes respectively, where $I_n=J_m$, results in an ($N+M-2$)\textsuperscript{th}-order tensor with entries $c_{i_1,\dots,i_{n-1},i_{n+1},\dots,i_N,j_1,\dots,j_{m-1},j_{m+1},\dots,j_M}= \\
    \sum_{i_n=1}^{I_n}a_{i_1,\dots,i_{n-1},i_n,i_{n+1},\dots,i_N}b_{j_1,\dots,j_{m-1},i_n,j_{m+1},\dots,j_M}$.

\noindent\textbf{Tensor inner product.}
    The \textit{inner product} of two $N$\textsuperscript{th}-order tensors, $\mathcalbf{A}\in\mathbb{R}^{I_1\times \dots \times I_N}$ and $\mathcalbf{B}\in\mathbb{R}^{I_1\times \dots \times I_N}$, results in a scalar, $c=\langle\mathcalbf{A},\mathcalbf{B}\rangle\in\mathbb{R}$, such that $c=\sum_{i_1,\dots,i_N}a_{i_1,\dots,i_N}b_{i_1,\dots,i_N}$.
    
\noindent\textbf{Rank-1 tensor.}
    An $N$\textsuperscript{th}-order tensor, $\mathcalbf{A}\in\mathbb{R}^{I_1\times \dots \times I_N}$, is said to be of rank-$R=1$, if it can be expressed as an outer product of $N$ vectors, $\mathcalbf{A} = \textbf{a}^{(1)} \circ \cdots \circ \textbf{a}^{(N)} = \bigcirc_{n=1}^N \textbf{a}^{(n)}$, where $\textbf{a}^{(n)} \in \mathbb{R}^{I_n}$.
    
\noindent\textbf{Rank-1 tensor inner product.}
    The \textit{inner product} between two $N$\textsuperscript{th}-order tensors, $\mathcalbf{A}\in\mathbb{R}^{I_1\times \dots \times I_N}$ and $\mathcalbf{B}\in\mathbb{R}^{I_1\times \dots \times I_N}$, can be expressed as a sequence of tensor contractions, $\langle\mathcalbf{A},\mathcalbf{B}\rangle = \mathcalbf{A} \times_1^1 \textbf{b}^{(1)} \times_2^1 \cdots \times_N^1 \textbf{b}^{(N)}$, if the tensor $\mathcalbf{B}$ is of rank-$1$, such that $\mathcalbf{B} = \bigcirc_{n=1}^N \textbf{b}^{(n)}$.
    
\noindent\textbf{Tensor-by-vector product.}
    A tensor-by-vector product is a special case of tensor contraction between an $N$\textsuperscript{th}-order tensor, $\mathcalbf{A}\in\mathbb{R}^{I_1\times \dots \times I_N}$, and an $1$\textsuperscript{st}-order tensor, $\mathbf{b}\in\mathbb{R}^{I_n}$, which results in a $(N-1)$-th order tensor, $\mathcalbf{C}\in\mathbb{R}^{I_1 \times \dots \times I_{n-1} \times I_{n+1} \times \cdots \times I_N}$, which is exactly 1 order smaller than $\mathcalbf{A}$. This operation is mathematically equivalent to performing the following steps in a sequence: (i) Matricize the tensor, $\textbf{A}_{(n)} \xleftarrow{unfold} \mathcalbf{A}$; (ii) Perform a linear combination (matrix-by-vector multiplication) in the $n$-th subspace, $\textbf{C}_{(n)} = \textbf{A}_{(n)}^T \textbf{b}$; (iii) Tensorize the result, $\mathcalbf{C} \xleftarrow{fold} \textbf{C}_{(n)}$. Therefore, the tensor-by-vector product can be interpreted as performing a linear combination in the $n$-th subspace.
    
\noindent\textbf{Tensor network diagram.}
    For visualization purpose and to gain physical insight into the multi-dimensional nature of tensor operations, we adopt the Tensor Network (TN) notation, whereby a tensor is diagrammatically represented as a circular vertex with outgoing edges; the number of edges equals the order of the tensor (see \cite{cichocki2016tensor} for more information on this graphical notation), as illustrated in Figure \ref{fig:basic_tensors}. If two vertices are connected by an edge, this connection represents a tensor contraction between two tensors over their shared common mode.

\begin{figure}[t]
    	\centering
    	\includegraphics[width=1.0\linewidth]{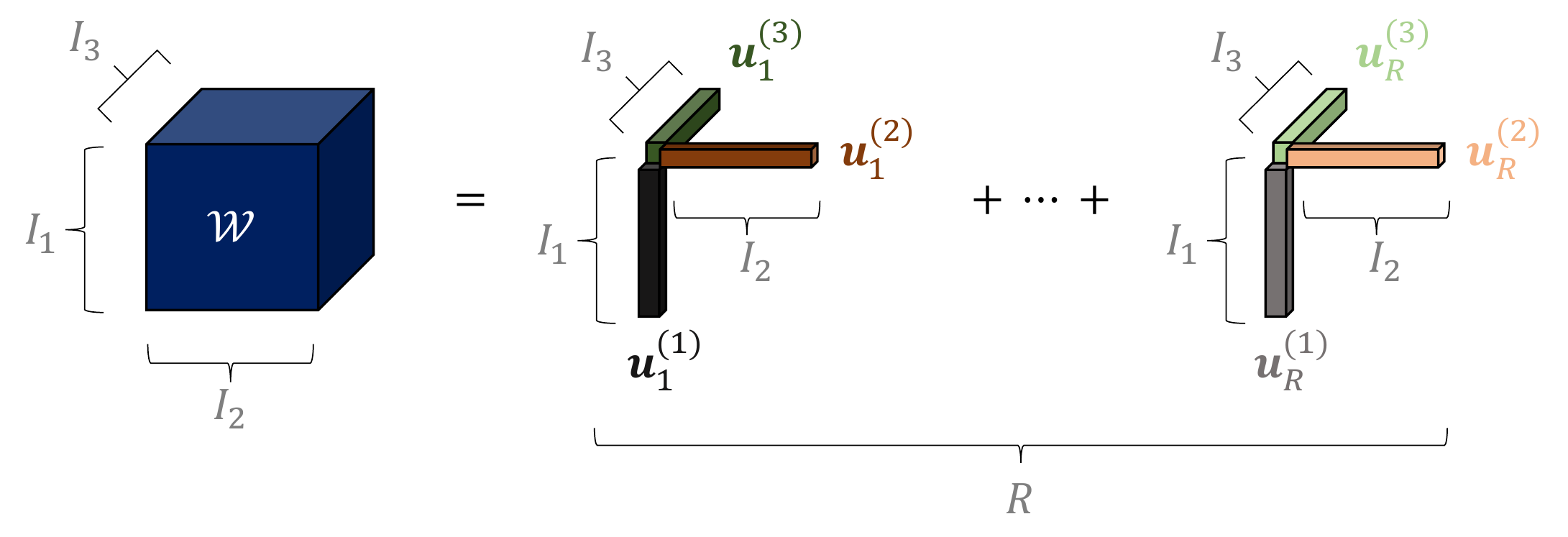}
    	\caption{Visualization of the Canonical Polyadic Decomposition (CPD) for an order-3 tensor, $\mathcalbf{W} \in \mathbb{R}^{I_1 \times I_2 \times I_3}$, according to (\ref{eq:cpd}). The sizes of tensor modes, $I_1$, $I_2$, and $I_3$, are labelled in gray.}
    	\label{fig:cpd_standard}
    \end{figure}

\noindent\textbf{Canonical Polyadic Decomposition. }
    Canonical Polyadic Decomposition (CPD) of rank, $R$, decomposes an order-$N$ tensor, $\mathcalbf{W} \in \mathbb{R}^{I_1 \times I_2 \times \cdots \times I_N}$, into a sum of outer products of $N$ vectors, and is given by
    \begin{equation} \label{eq:cpd}
        \begin{aligned}
            \mathcalbf{W} 
            &= \sum_{r=1}^R \textbf{u}_r^{(1)} \circ \textbf{u}_r^{(2)} \circ \cdots \circ \textbf{u}_r^{(N)} \\
    \end{aligned}
    \end{equation}
    where $\textbf{u}_r^{(n)} \in \mathbb{R}^{I_n}$ denotes the $r$-th mode-$n$ factor vector, as illustrated diagrammatically in Figure \ref{fig:cpd_standard}.
 
\begin{figure}[t]
    	\centering
    	
    	\begin{subfigure}[b]{\columnwidth}
    	\includegraphics[width=1.0\linewidth]{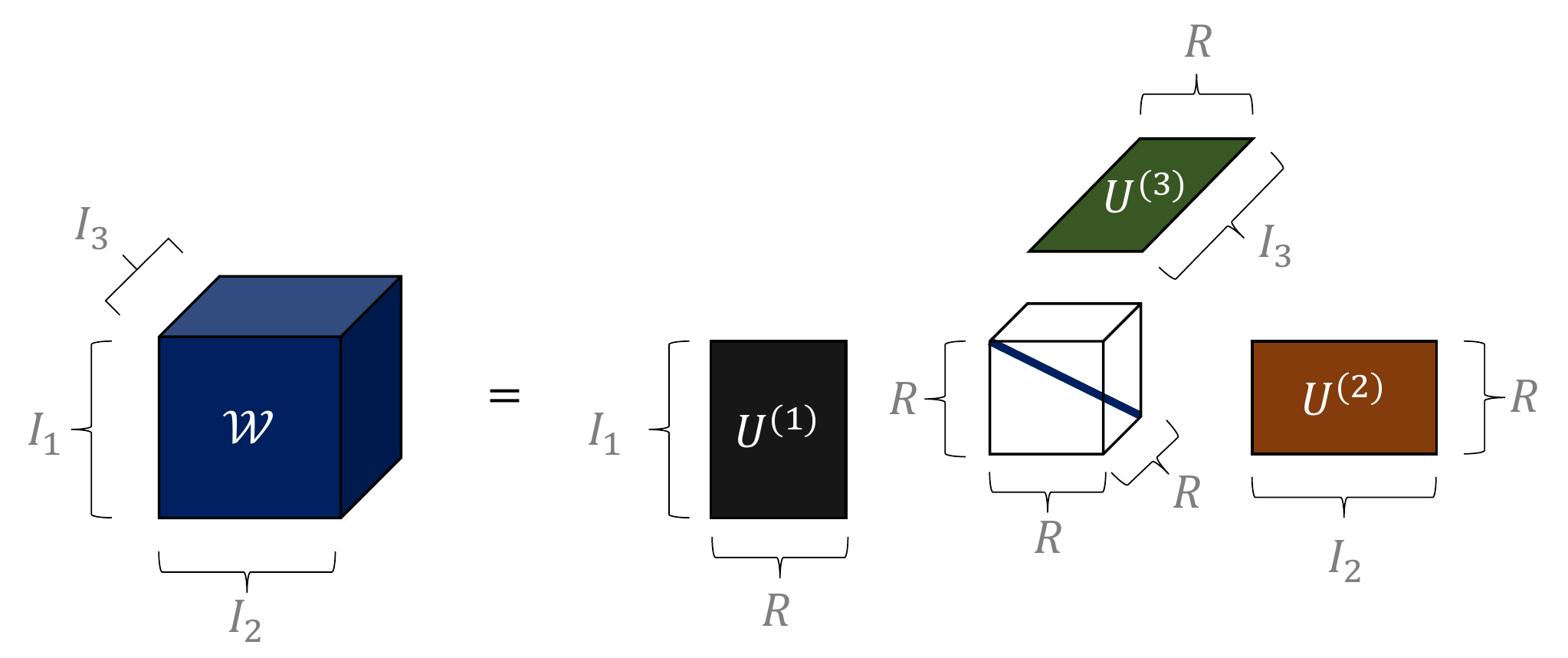}
    	\end{subfigure}
    	
    	\vspace{10mm}
    	
        \begin{subfigure}[b]{\columnwidth}
    	\includegraphics[width=1.0\linewidth]{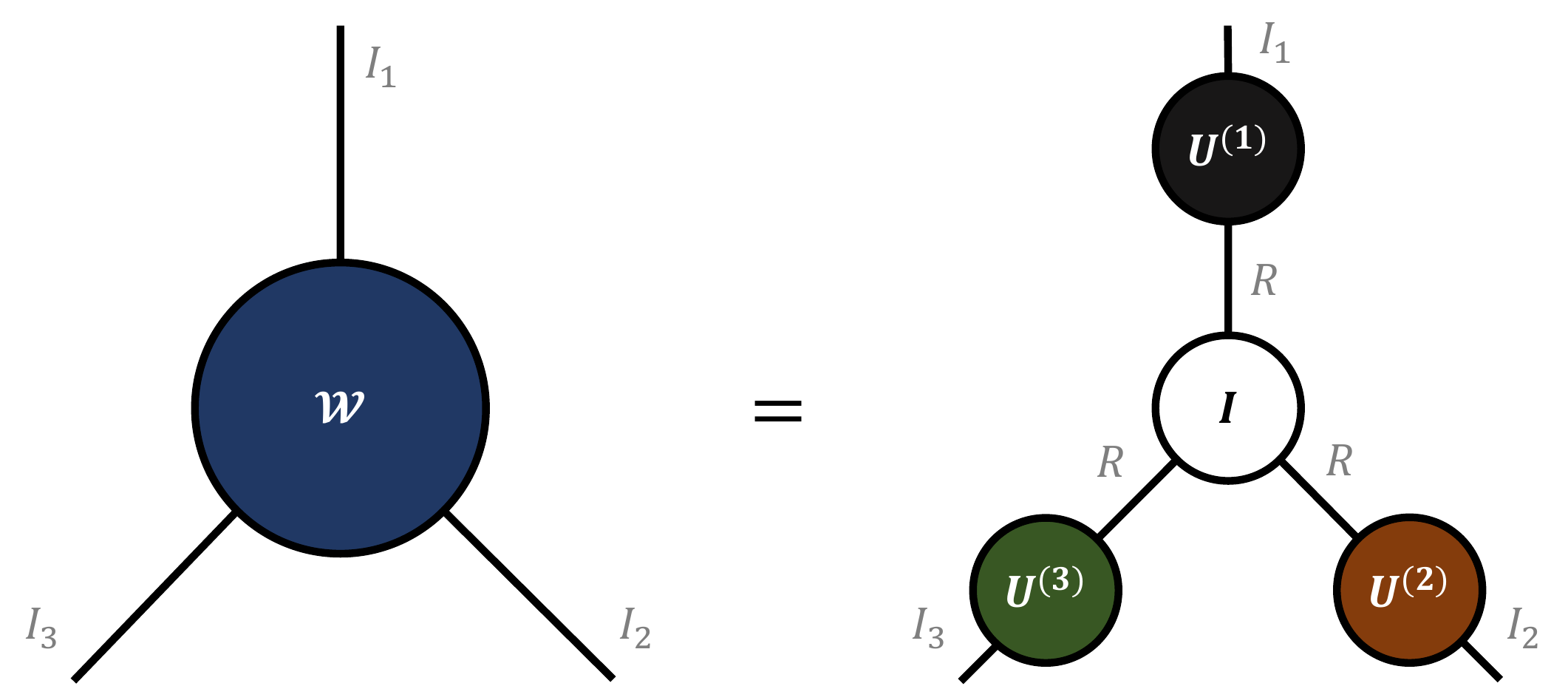}
    	
        \end{subfigure}
        
    	\caption{Alternative visualization of the Canonical Polyadic Decomposition for an order-3 tensor, $\mathcalbf{W} \in \mathbb{R}^{I_1 \times I_2 \times I_3}$, according to (\ref{eq:cpd_contraction}) in: (Top) standard matrix notation and (Bottom) TN notation. The sizes of tensor modes, $I_1$, $I_2$, and $I_3$, are labelled in gray.}
    	\label{fig:cpd}
    	
    \end{figure}

    Alternatively, for each mode-$n$, we can stack the vectors, $\textbf{u}_r^{(n)}$ for $r=1, \ldots, R$, to form the factor matrix, $\textbf{U} \in \mathbb{R}^{I_n \times R}$, as $\textbf{U} = [\textbf{u}_1^{(n)}, \textbf{u}_2^{(n)}, \ldots, \textbf{u}_R^{(n)}]$. This allows us to express the CPD as a series of tensor contractions, given by
    \begin{equation} \label{eq:cpd_contraction}
        \begin{aligned}
            \mathcalbf{W} 
            &= \mathcalbf{I} \times_1^1 \textbf{U}^{(1)} \times_2^1 \textbf{U}^{(2)} \times_3^1 \cdots \times_N^1 \textbf{U}^{(N)} \\
        \end{aligned}
    \end{equation}
    as illustrated in Figure \ref{fig:cpd} in both standard matrix notation and TN notation.

    

    The mode-$n$ matricization of a tensor in its CPD format, $\mathbf{W}_{(n)} \in \mathbb{R}^{I_n \times I_1 \cdots I_{n-1} I_{n+1} \cdots I_N}$, can be expressed as a series of Khatri-Rao products of the factors matrices, as
    \begin{equation}
        \begin{aligned}
            \mathbf{W}_{(n)}
            &= \textbf{U}^{(n)} \left( \textbf{U}^{(N)} \odot \cdots \odot \textbf{U}^{(n+1)} \odot \textbf{U}^{(n-1)} \odot \textbf{U}^{(1)} \right)^T \\
            &= \textbf{U}^{(n)} \left( \bigodot_{i=N, i \neq n}^1 \textbf{U}^{(i)} \right)^T \\
            &= \textbf{U}^{(n)} \textbf{U}^{(-n)^T}
        \end{aligned}
    \end{equation}
    where $\textbf{U}^{(-n)} = \left( \bigodot_{i=N, i \neq n}^1 \textbf{U}^{(i)} \right)$. 
    
    Similarly, the vectorization of a tensor in its CPD format, $\vectorize(\mathcalbf{W}) \in \mathbb{R}^{I_1 \cdots I_N}$, can be expressed as 
    \begin{equation}
        \begin{aligned}
        \vectorize(\mathcalbf{W})
        &= \left( \textbf{U}^{(N)} \odot \textbf{U}^{(N-1)} \odot \cdots \odot \textbf{U}^{(1)} \right) \textbf{1} \\
        &= \left( \bigodot_{i=N}^1 \textbf{U}^{(i)} \right) \textbf{1} \\
        \end{aligned}
    \end{equation}
    where $\textbf{1}$ denotes a vector of unities.

\subsection{Graphs and Signals on Graphs} \label{sec:graph_sig}

\begin{figure}[h]
	\centering
	\includegraphics[width=1.0\linewidth]{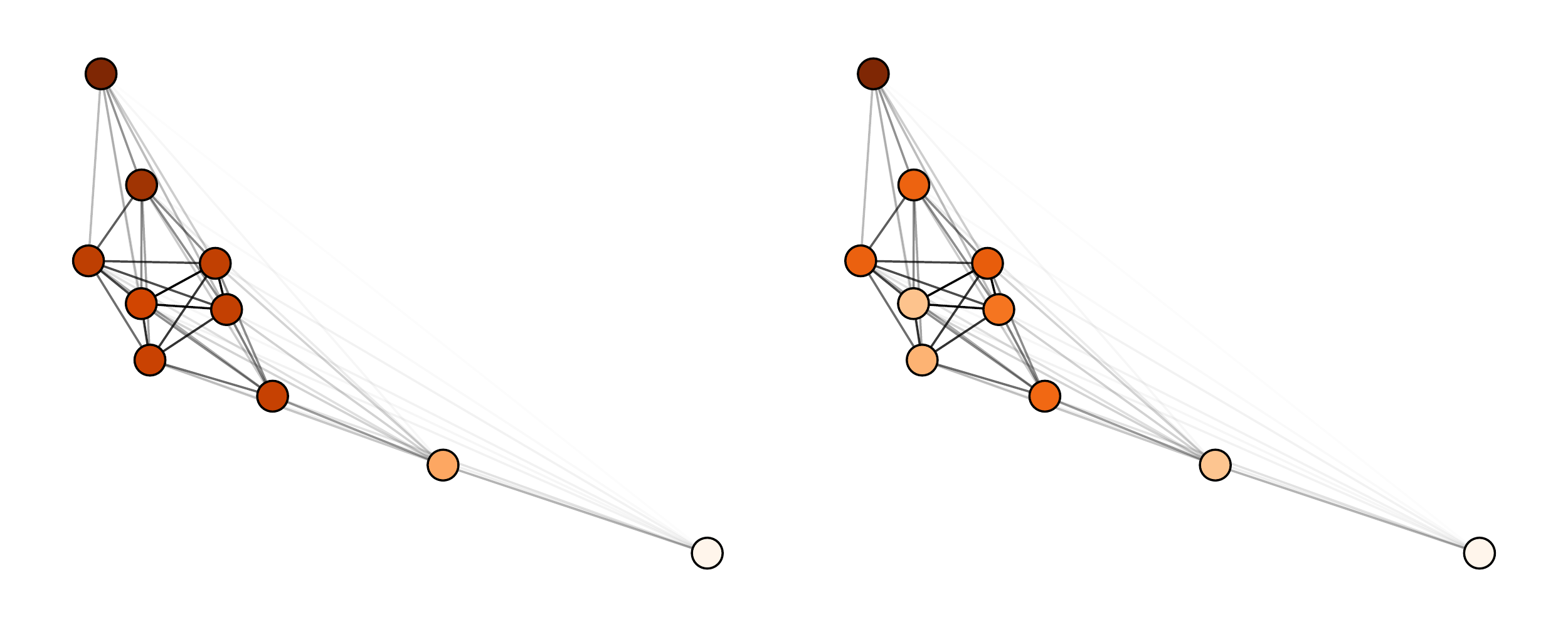}
	\caption{Smooth (left) and non-smooth (right) graph signals. The vertices of the illustrated graphs are coloured proportionally to the graph signal values, while the edges are also coloured proportionally to their edge weight. A graph signal is considered smooth if it varies slowly over closely connected neighbouring vertices.}
	\label{fig:smooth_graph_sig}
\end{figure}

A graph, $\mathcal{G} = \{\mathcal{V}, \mathcal{E}\}$, is defined by a set of $I$ vertices (or nodes) $\textit{v}_i \subset \mathcal{V}$ for $i = 1, \ldots , I$, and a set of edges connecting the ${i_1}^{th}$ and ${i_2}^{th}$ vertex, $\textit{e}_{i_1, i_2} = (\textit{v}_{i_1}, \textit{v}_{i_2}) \in \mathcal{E}$, for $i_1=1,\ldots,I$ and $i_2=1,\ldots,I$. A graph can be fully described by the weighted adjacency matrix, $\textbf{A} \in \mathbb{R} ^ {I \times I}$, such that $\textit{a}_{i_1, i_2} > 0$ if $\textit{e}_{i_1, i_2} \in \mathcal{E}$, and $\textit{a}_{i_1, i_2} = 0$ if $\textit{e}_{i_1, i_2} \notin \mathcal{E}$. Alternatively, a graph can be described in terms of its Laplacian matrix, $\mathbf{L} \in \mathbb{R} ^ {I \times I}$, defined as $\textbf{L} = \textbf{D} - \textbf{A}$, where $\textbf{D} \in \mathbb{R} ^ {I \times I}$ is the diagonal degree matrix such that $d_{i_1, i_1} = \sum_{i_2} \textit{a}_{i_1, i_2}$ \cite{stankovic2019graph}. 



A graph signal can be represented by a vector, $\textbf{u} \in \mathbb{R} ^ {I}$, which associates a scalar value to each of its $I$ vertices. The smoothness of such a graph signal with respect to the underlying graph topology can then be quantified as $s=\textbf{u}^T \textbf{L} \textbf{u}$, which measures the change in signal value between neighbouring (i.e., connected) vertices \cite{stankovic2019graphII}. Therefore, a graph signal is said to be smooth if it does not change abruptly across connected nodes, as illustrated in Figure \ref{fig:smooth_graph_sig}. 

For a given set of $R$ different graph signals, the total smoothness over all such signals can be computed as $s=\sum_{r} \textbf{u}_r^T \textbf{L} \textbf{u}_r$. Equivalently, the given set of $R$ graph signals can be stacked as a set of column vectors into a matrix, $\textbf{U} \in \mathbb{R}^{I \times R}$. The smoothness can then be computed as 
\begin{equation} \label{eq:graph_sig_smoothness}
    \begin{aligned}
        s 
        = \sum_{r} \textbf{u}_r^T \textbf{L} \textbf{u}_r 
        = \tr \left( \textbf{U}^T \textbf{L} \textbf{U} \right) 
    \end{aligned}
\end{equation}


\section{Graph Regularized Tensor Regression} \label{sec:grtr}

\subsection{Tensor Regression}

For a set of tensor-structured inputs, $\mathcalbf{X} \in \mathbb{R}^{I_1 \times I_2 \times \cdots \times I_N}$, and a set of corresponding output labels,  $y \in \mathbb{R}$, the tensor regression \cite{5986711} can be expressed as 

\begin{equation} \label{eq:basic_tensor_regression}
    y = \langle \mathcalbf{W}, \mathcalbf{X} \rangle + b
\end{equation}

Analogously to standard regression, Tensor Regression (TR) aims to learn: (i) the regression constant, $b \in \mathbb{R}$, and (ii) the weight tensor, $\mathcalbf{W} \in \mathbb{R}^{I_1 \times I_2 \times \cdots \times I_N}$, in the CPD format, such that the prediction mean-squared-error (MSE) in (\ref{eq:basic_tensor_regression}) is minimized for a given dataset. 

\begin{figure} 
	\centering
	\includegraphics[width=1.0\linewidth]{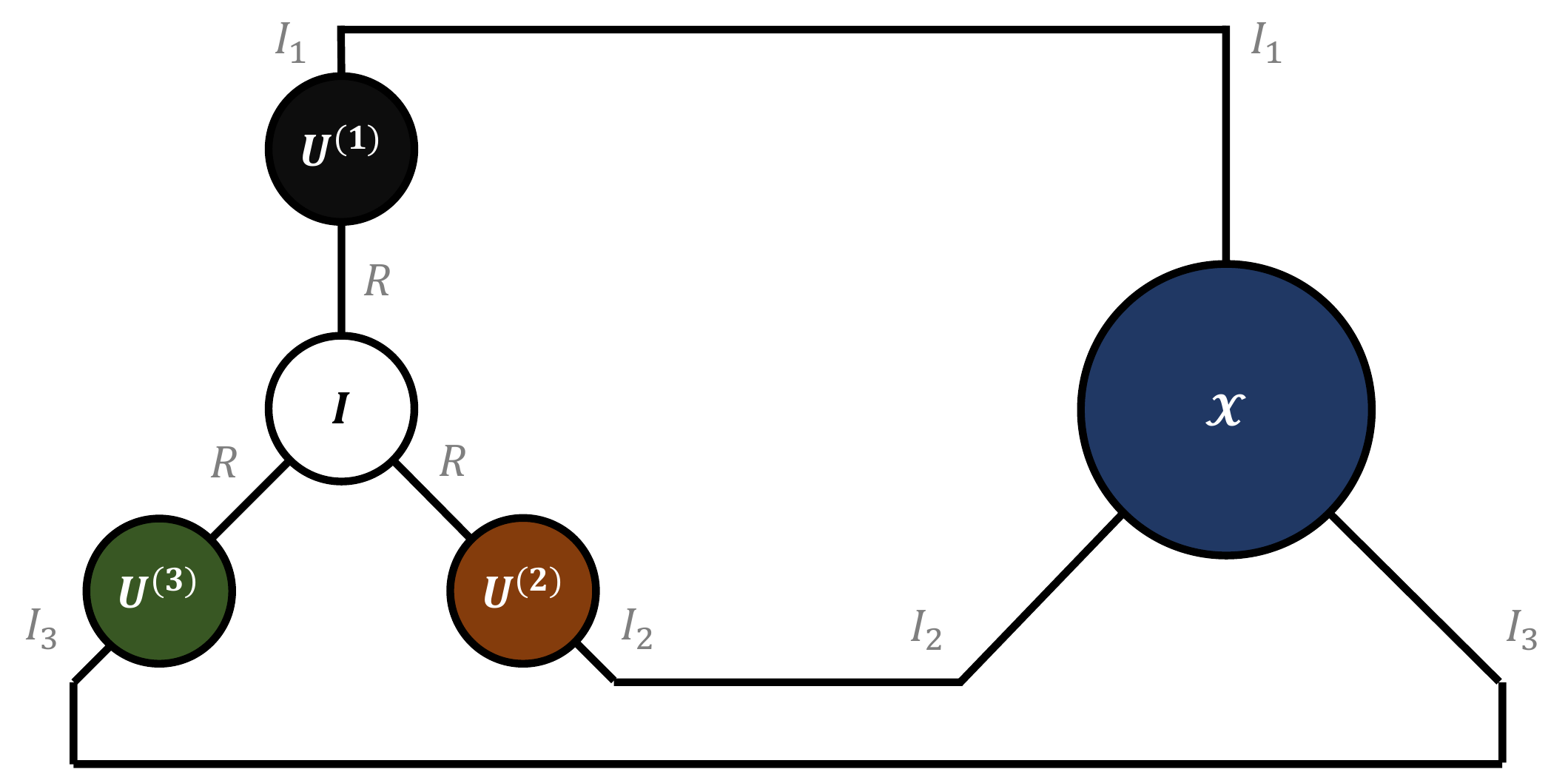}
	\caption{Tensor network diagram of the inner product, $\langle \mathcalbf{W}, \mathcalbf{X} \rangle$, within an order-3 tensor regression. The weight tensor, $\mathcalbf{W} \in \mathbb{R}^{I_1 \times I_2 \times \cdots \times I_N}$, on the left is factorized by the CPD according to equation (\ref{eq:cpd_contraction}). The connected tensor modes represent tensor contractions along these modes.}
	\label{fig:tensor_regression}
\end{figure}

\begin{remark}
Instead of directly learning the large weight tensor, $\mathcalbf{W}$, tensor regression learns its low-rank CPD factor matrices, $\textbf{U}^{(n)}$ (one for each of the $N$ input dimensions), such that equation (\ref{eq:cpd_contraction})
holds. The tensor regression framework in (\ref{eq:basic_tensor_regression}) is illustrated in Figure \ref{fig:tensor_regression} in the TN notation.
\end{remark}

The learning of the CPD factor matrices is often achieved through the Alternating Least Squares (ALS) procedure \cite{5986711}, which iteratively updates each factor matrix by minimizing the MSE with respect to mode-$n$, $n=1,\ldots,N$, until convergence. 

\begin{remark}
For a learned weight tensor, $\mathcalbf{W}$, the tensor regression framework in (\ref{eq:basic_tensor_regression}) is mathematically equivalent to a classical linear regression, $y = \textbf{w}^T\textbf{x} + b$, where $\textbf{w}$ and $\textbf{x}$ are respectively the vectorized versions of $\mathcalbf{W}$ and $\mathcalbf{X}$, given by 
\begin{equation} \label{eq:lr_equal_tr}
    \begin{aligned}
        y 
        = \langle \mathcalbf{X}, \mathcalbf{W} \rangle + b 
        = \vectorize (\mathcalbf{X})^T \vectorize (\mathcalbf{W}) + b 
        = \textbf{x}^T \textbf{w} + b 
    \end{aligned}
\end{equation}
The main difference (and advantage) of the tensor regression framework over a standard multivariate regression approach is that it maintains the multi-dimensional structure of the original data, allowing us to exploit its multi-way, low-rank CPD structure during the training (i.e., parameter estimation) stage. This reduces both the computational complexity and the likelihood of over-fitting, while maintaining interpretability.
\end{remark}

\subsection{Graph Regularized Tensor Regression}

In practice, when learning the factor matrices, $\textbf{U}^{(n)} \in \mathbb{R}^{I_n \times R}$, in addition to minimizing the MSE, the parameter estimation is often regularized by the $L_2$-norm of the factor matrices. However, such a regularization scheme is ignorant to the intra-modal relations present in the data. In other words, given some domain knowledge about the relational structure within mode-$n$ (e.g., sector relations among $I_n$ stocks), a naive $L_2$-regularization scheme cannot incorporate such domain knowledge into the model training process. 

To this end, we propose an extension to the classical tensor regression paradigm to enable the incorporation of domain knowledge into the learning process. Given a graph Laplacian matrix, $\textbf{L}^{(n)} \in \mathbb{R}^{I_n \times I_n}$, which encodes domain knowledge about the relational structure within mode-$n$, we aid the learning of the mode-$n$ factor matrix, $\textbf{U}^{(n)} \in \mathbb{R}^{I_n \times R}$, through a corresponding mode-$n$ graph Laplacian regularization, as discussed in Section \ref{sec:graph_sig}. In this way, domain-knowledge about data modes informs the learning, which both reduces the likelihood of over-fitting, and accelerates the learning process. 

\begin{figure*}
	\centering
	\includegraphics[width=1.0\linewidth]{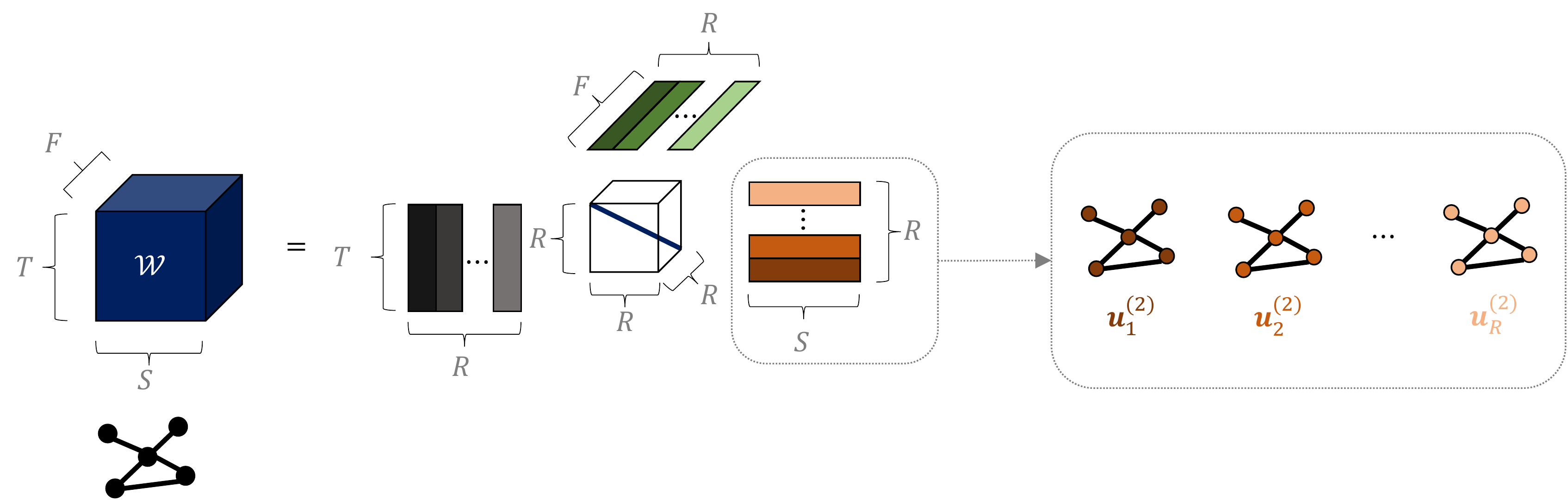}
	\caption{Principle of graph-regularized tensor regression. The order-3 weight tensor, $\mathcalbf{W} \in \mathbb{R}^{T \times S \times F}$, is represented as a rank-$R$ CPD tensor decomposition, where the corresponding factor matrices are of dimensions, $\textbf{U}^{(1)} \in \mathbb{R}^{T \times R}$, $\textbf{U}^{(2)} \in \mathbb{R}^{S \times R}$, and $\textbf{U}^{(3)} \in \mathbb{R}^{F \times R}$.  Our aim is to regularize the learning of the mode-$2$ factor matrix, $\textbf{U}^{(2)} \in \mathbb{R}^{S \times R}$, such that it reflects the sector information of the $S$ stocks as domain knowledge. To this end,  we construct a graph with $S$ vertices representing $S$ stocks, where an edge connects a pair of vertices if they belong to the same sector.  Given such a mode-$2$ graph, the corresponding factor matrix, $\textbf{U} = [\textbf{u}_1^{(n)}, \textbf{u}_2^{(n)}, \ldots, \textbf{u}_R^{(n)}]$, can be viewed as a set of $R$ graph signals (its column vectors), which can be learned so as to be smooth with respect to the underlying stocks graph. This makes it possible to regularize the learning of the factor matrix and results, as desired, in similar weights for stocks from the same sector.}
	\label{fig:weights_on_graphs}
\end{figure*}

\begin{remark}
The column vectors of the mode-$n$ CPD factor matrix, $\textbf{u}^{(n)}_r \in \mathbb{R}^{I_n}$ for $r=1,\ldots,R$, can be viewed as graph signals. In this way, the estimation of the factor matrix can be interpreted as the estimation of $R$ graph signals that are smooth with respect to the graph topology encoded by the mode-$n$ Laplacian matrix, $\textbf{L}^{(n)} \in \mathbb{R}^{I_n \times I_n}$.
\end{remark}


\begin{example} \label{ex:row_vs_column}
Consider an order-$3$ input tensor, $\mathcalbf{X} \in \mathbb{R}^{T \times S \times F}$, where $F$ features for $S$ stocks are indexed over $T$ time-steps. For this data, we wish to learn a tensor regression model, such that the mode-2 (the stocks mode) factor matrix, $\textbf{U}^{(2)} \in \mathbb{R}^{S \times R}$, incorporates the sector knowledge of the $S$ stocks, as illustrated in Figure \ref{fig:weights_on_graphs}.  To this end, we construct a graph with $S$ vertices representing $S$ different stocks, where the edges connect a pair of stocks if they belong to the same sector. Given this graph, we can interpret the column vectors of the mode-$2$ factor matrix, $\textbf{u}_1^{(n)}, \textbf{u}_2^{(n)}, \ldots, \textbf{u}_R^{(n)}$, as a set of $R$ graph signals. Therefore, the learning of the factor matrix, $\textbf{U}^{(2)}$, can be regularised by ensuring that its column vectors, $\textbf{u}^{(2)}_r$, $r=1,\ldots,R$, are smooth with respect to the underlying stock graph (i.e., the graph signal smoothness, $s=\tr \left( \textbf{U}^{(2)^T} \textbf{L}^{(2)} \textbf{U}^{(2)} \right)$, is small). 
\end{example}



To formally introduce the concept of graph regularization in the classical tensor regression framework, we define a mode-$n$ graph Laplacian matrix, $\textbf{L}^{(n)}$, for each of the $N$ tensor modes, and assign a corresponding regularization constant, $\lambda^{(n)}$. The mode-wise graph Laplacian regularization terms are then included into the MSE loss function, calculated over $m=1, \ldots, M$ samples, to yield the following loss function to be minimized


\begin{equation} \label{eq:grtr_loss}
    \begin{aligned}
        \mathcal{L} = 
        & \underbrace{\frac{1}{M} \sum_{m=1}^M \frac{1}{2} \left( y^{(m)} - \Big \langle \mathcalbf{W}, \mathcalbf{X}^{(m)} \Big \rangle - b \right)^2}_{\text{Error term}} \\
        & + \underbrace{\frac{1}{N} \sum_{n=1}^N \frac{1}{2} \left( \lambda^{(n)} \tr \left( \textbf{U}^{(n)^T} \textbf{L}^{(n)} \textbf{U}^{(n)} \right) \right)}_{\text{Regularization term}}
    \end{aligned}
\end{equation}

The constant, $b \in \mathbb{R}$, that minimizes the loss function, $\mathcal{L}$, is estimated based on the derivative of the loss function with respect to this constant using the chain-rule, resulting in

\begin{equation} \label{eq:partial_derivative_constant}
    \begin{aligned}
        \frac{\partial \mathcal{L}}{\partial b} = 
        & - \frac{1}{M} \sum_{m=1}^M \left( y^{(m)} - \Big \langle \mathcalbf{W}, \mathcalbf{X}^{(m)} \Big \rangle - b \right) 
    \end{aligned}
\end{equation}


The factor matrices that minimize the loss function, are estimated by the Alternating Least Squares (ALS) framework, which iteratively minimizes the loss function sequentially one-mode-at-a-time. To this end, we can apply the properties discussed in Section \ref{sec:prelim_tensors} and matricize the tensor contraction with respect to the $n$-th mode as


\begin{equation}
    \begin{aligned}
        \Big \langle \mathcalbf{W}, \mathcalbf{X}^{(m)} \Big \rangle
        &= \Big \langle \mathbf{W}_{(n)}, \mathbf{X}^{(m)}_{(n)} \Big \rangle \\
        &= \Big \langle \textbf{U}^{(n)} \textbf{U}^{(-n)^T}, \mathbf{X}^{(m)}_{(n)} \Big \rangle \\
        &= \tr \left( \textbf{U}^{(n)} \textbf{U}^{(-n)^T} \mathbf{X}^{(m)^T}_{(n)} \right) \\
    \end{aligned}
\end{equation}
where $\textbf{U}^{(-n)} = \left( \bigodot_{i=N, i \neq n}^1 \textbf{U}^{(i)} \right)$.

This mode-$n$ matricized formulation allows for the computation of the partial derivative with respect to mode-$n$ factor matrix, by leveraging on the well-known matrix derivatives rules 
\begin{equation}
    \begin{aligned}
        \frac{\partial}{\partial \textbf{U}^{(n)}} \tr \left( \textbf{U}^{(n)} \textbf{U}^{(-n)^T} \mathbf{X}^{(m)^T}_{(n)} \right) =  \mathbf{X}^{(m)}_{(n)} \textbf{U}^{(-n)} 
    \end{aligned}
\end{equation}

Through the chain rule, the above results can be included into the calculation of the derivative of the loss function with respect to the mode-$n$ factor matrix as


\begin{equation} \label{eq:partial_derivative_factors}
    \begin{aligned}
        \frac{\partial \mathcal{L}}{\partial \textbf{U}^{(n)}} = 
        & - \frac{1}{M} \sum_{m=1}^M \epsilon^{(m)} \mathbf{X}^{(m)}_{(n)} \textbf{U}^{(-n)}  \\
        & + \frac{1}{N} \left( \lambda^{(n)} \textbf{L}^{(n)} \textbf{U}^{(n)} \right)
    \end{aligned}
\end{equation}
where $\epsilon^{(m)} = \left( y^{(m)} - \Big \langle \mathbf{W}_{(n)}, \mathbf{X}^{(m)}_{(n)} \Big \rangle - b \right)$ denotes the prediction residual corresponding to the $m$-th sample.

\begin{remark}
Equation \ref{eq:partial_derivative_factors} considers the general case where each mode of the input tensor, $\mathcalbf{X} \in \mathbb{R}^{I_1 \times I_2 \times \cdots \times I_N}$, is associated with an underlying graph domain. However, if the $n$-th mode does not have an associated Laplacian matrix, then the corresponding regularization parameter, $\lambda^{(n)}$, can simply be set to $0$ to ignore the regularization term in the loss function.
\end{remark}

Based on the partial derivatives in (\ref{eq:partial_derivative_factors}), a gradient-descent optimization framework can be employed to iteratively update the factor matrices until convergence. The learning algorithm is summarized in Algorithm \ref{alg:grtr_als}. 
\begin{algorithm}
\SetAlgoLined

    \textbf{Input:} Set of features and labels, $\{ \mathcalbf{X}^{(m)}, y^{(m)} \}$, set of Laplacian matrices, $\{ \textbf{L}^{(n)} \}$, set of regularization constants, $\{ {\lambda^{(n)}, \rho^{(n)}} \}$, model rank, $R$, learning rate, $\alpha$, model error tolerance, $t$, maximum steps $K$.
    
    \textbf{Output:} Tensor weights, $\textbf{U}^{(1)}, \ldots, \textbf{U}^{(N)}$
    
    \textbf{}\\
    
    Initialise $\textbf{U}^{(n)} \in \mathbb{R}^{I_n \times R}$ randomly, $n=1, \ldots, N$. 
    
    Initialise $b \in \mathbb{R}$ randomly.
    
    Initialise $\epsilon = \infty$.
    
    Initialise $k=0$.
    
    \textbf{} \\
    
    \While {$\epsilon > t$ and $k<K$}{
        $k \gets k+1$
        
        $\mathcalbf{W} \gets \mathcalbf{I} \times_1^1 \textbf{U}^{(1)} \times_2^1 \textbf{U}^{(2)} \times_3^1 \cdots \times_N^1 \textbf{U}^{(N)}$
        
        $\hat{y}^{(m)} \gets \Big \langle \mathcalbf{W}, \mathcalbf{X}^{(m)} \Big \rangle + b$
        
        $\epsilon \gets \frac{1}{M} \sum_{m=1}^M ({y}^{(m)} - \hat{y}^{(m)})^2$
        
        \For {$n=1, \ldots, N$}{
            Compute $\frac{\partial \mathcal{L}}{\partial \textbf{U}^{(n)}}$ according to \ref{eq:partial_derivative_constant}
            
            Compute $\frac{\partial \mathcal{L}}{\partial b}$ according to \ref{eq:partial_derivative_factors}
            
            Update $\textbf{U}^{(n)} \gets \textbf{U}^{(n)} - \alpha \frac{\partial \mathcal{L}}{\partial \textbf{U}^{(n)}}$
            
            Update $b \gets b - \alpha \frac{\partial \mathcal{L}}{\partial b}$
        }
    }
    
    \textbf{Return:} $\textbf{U}^{(1)}, \ldots, \textbf{U}^{(N)}$
    
    \caption{GRTR-ALS-GD} \label{alg:grtr_als}
\end{algorithm}



\begin{figure*}[t]
	\centering
	\includegraphics[width=1.0\linewidth]{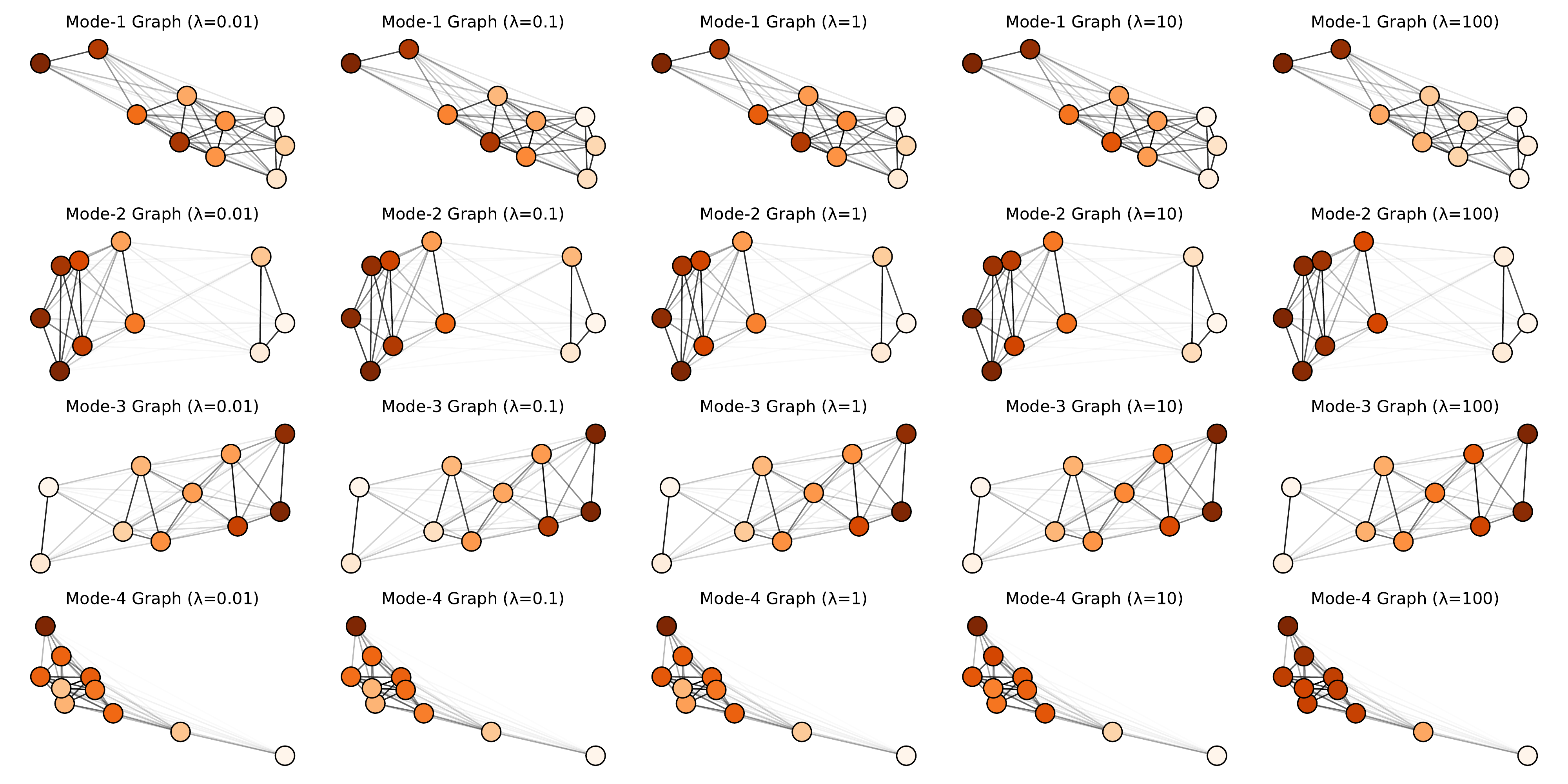}
	\caption{Effect of the graph regularization parameter, $\lambda^{(n)}$, on the estimated factor vectors, $\textbf{u}^{(n)} \in \mathbb{R}^{I_n}$. The row panels illustrate $N=4$ different graph domains, each with $I_n=10$ vertices that are coloured proportionally to the values of the corresponding factor vectors. The column panels illustrate the factor vectors for different values of $\lambda^{(n)}$. Notice how the factor vectors become smoother with respect to the underlying graph with an increase in regularization parameter, as neighbouring vertices exhibit increasingly similar colours with larger $\lambda^{(n)}$.}
	\label{fig:regularization_effects}
\end{figure*}

\subsection{Choice of Model Rank}

A key hyper-parameter choice within the GRTR model is the model rank. A larger rank increases the expressive power of the model, but it also increases the space complexity of the model which can lead to over-fitting. In practice, the optimal choice of model rank can be determined via grid-search, whereby the rank with the smallest cross-validation error is used for the final model. 

\subsection{Choice of Regularization Parameter}

The regularization parameter, $\lambda^{(n)}$, is another important hyper-parameter. As $\lambda^{(n)}$ increases, the regularization term begins to dominate the loss function over the error term in (\ref{eq:grtr_loss}), which reduces the likelihood of in-sample over-fitting and results in better out-of-sample performance. In practice, the optimal choice of regularization parameter can also be determined empirically via grid search.

\begin{example}
To illustrate the effect of the graph regularization parameter, consider a GRTR model of rank $R=1$ , whereby the mode-$n$ factor matrix, $\textbf{U}^{(n)} \in \mathbb{R}^{I_n \times R}$, is reduced to a column vector, $\textbf{u}^{(n)} \in \mathbb{R}^{I_n}$. This factor vector is effectively a graph signal, which contains a scalar value for each of the $I_n$ vertices in the mode-$n$ graph implied by the corresponding Laplacian matrix, $\textbf{L}^{(n)} \in \mathbb{R}^{I_n \times I_n}$. For large values of $\lambda^{(n)}$, this leads to highly smooth vectors, $\textbf{u}^{(n)} \in \mathbb{R}^{I_n}$, where the difference between entries, $|u^{(n)}_i - u^{(n)}_j|$, is small if the $i$-th and $j$-th vertices are closely connected in the underlying mode-$n$ graph. This effect is illustrated in Figure \ref{fig:regularization_effects}. 
\end{example}

\subsection{Complexity Analysis}

We now proceed to analyse the space complexity (i.e., the number of trainable parameters) and the time complexity (i.e., the number of basic computations) of the proposed algorithm. 

Once the learning is complete, inference can be performed by reconstructing the weight tensor, $\mathcalbf{W}$, according to (\ref{eq:cpd}), which requires a space complexity of $\mathcal{O} \left( I^N \right)$, where $I_n = I$ for all $n$. In this way, the time complexity of the inference becomes the same as in standard linear regression, $\mathcal{O} \left( I^N \right)$. This is because tensor regression can be re-written in a vectorized format, as shown in (\ref{eq:lr_equal_tr}).

Alternatively, the trained factor matrices, $\textbf{U}^{(1)}, \ldots, \textbf{U}^{(N)}$,  can be kept in a low-rank CPD format, without reconstructing the full weight tensor, $\mathcalbf{W}$. This reduces the space-complexity from an exponential one, $\mathcal{O} \left( I^N \right)$, to a linear one, $\mathcal{O} \left( NIR \right)$, in the tensor order, $N$. However, in this way the time-complexity of inference would increase to $\mathcal{O} \left( RI^N \right)$, as this requires evaluation of a series of tensor contractions
\begin{equation}
    \begin{aligned}
        y 
        &= \langle \mathcalbf{X}, \mathcalbf{W} \rangle + b \\
        &= \mathcalbf{X} \times_{1, \ldots, N}^{1, \ldots, N} \left( \mathcalbf{I} \times_1^1 \textbf{U}^{(1)} \times_2^1 \cdots \times_N^1 \textbf{U}^{(N)} \right) + b \\
        &= \left( \mathcalbf{X} \times_1^1 \textbf{U}^{(1)} \times_2^1 \cdots \times_N^1 \textbf{U}^{(N)} \right) \times_{1, \ldots, N}^{1, \ldots, N} \mathcalbf{I} + b \\
    \end{aligned}
\end{equation}

\section{Experiments on Synthetic Data} \label{sec:exp_syn}

To validate the ability of the proposed model to capture inter- and intra-modal relations in data, as opposed to classical models, we conducted an experiment on synthetic data involving large and noisy tensor data. For reproducibility, all experiment code can be found on GitHub\footnote{\url{www.github.com/gylx/tensor\_regression\_graph}}. 

\subsection{Experiment Setting}

Consider a regression problem which is fully determined by an unknown CPD weight tensor, $\mathcalbf{W} \in \mathbb{R}^{I_1 \times \cdots \times I_N}$. This system generates a set of $M$ observations, $\left( \mathcalbf{X}^{(m)}, y^{(m)} \right)$, such that $y^{(m)} = \langle \mathcalbf{W}, \mathcalbf{X}^{(m)} \rangle + \eta^{(m)}$, where $\eta$ is additive random noise. Given this set of noisy observations, the goal of the experiment is to estimate the ground-truth weight tensor, $\Tilde{\mathcalbf{W}} \in \mathbb{R}^{I_1 \times \cdots \times I_N}$, using regression methods, such that the Mean Squared Error (MSE), $ \norm{\Tilde{\mathcalbf{W}} - {\mathcalbf{W}}}^2_2$, is as small as possible.

\subsection{Data}

To carry out the proposed regression experiment, $N=4$ ground-truth factor matrices, $\textbf{U}^{(n)} \in \mathbb{R}^{10 \times 5}$, were sampled randomly from a uniform distribution with rank $R=5$ and dimensions $I_n=10$, $n=1,\ldots,4$. The factor matrices were then contracted according to (\ref{eq:cpd_contraction}) to generate the ground-truth weight tensor of order-$4$, $\mathcalbf{W} \in \mathbb{R}^{10 \times 10 \times 10 \times 10}$. A set of $M=125$ input tensors, $\mathcalbf{X}^{(m)} \in \mathbb{R}^{10 \times 10 \times 10 \times 10}$, were then sampled randomly from a Gaussian distribution with zero-mean and unit-variance, which were then contracted with the weight tensor to generate a corresponding set of regression labels, $y^{(m)} = \langle \mathcalbf{W}, \mathcalbf{X}^{(m)} \rangle + \eta^{(m)}$, where $\eta^{(m)}$ is additive noise sampled from a Gaussian distribution with zero-mean and variance, $\sigma_{\eta}^2$, such that $\sigma_{\eta} = 0.5 \sigma_{y}$. The $M=125$ observations were then split according to a standard $80\%-20\%$ train-test split.

\begin{remark}
The input tensors, $\mathcalbf{X}^{(m)} \in \mathbb{R}^{10 \times 10 \times 10 \times 10}$, are order-$4$ tensors with modal dimensions $I_n=10$, giving effectively $10^4=10,000$ features for every sample. However, only $100$ samples were available for the training set, which is $2$ orders of magnitude smaller than the number of features. The Curse-of-Dimensionality inherent to this experiment, coupled with a low signal-to-noise ratio of $\frac{\sigma_{y}}{\sigma_{\eta}}=2$, is ubiquitous in financial Big Data. This poses significant challenges to traditional models, both in terms of complexity and over-fitting. 
\end{remark}

\subsection{Incorporation of Domain Knowledge through Graphs}

The proposed GRTR framework allows for the incorporation of domain knowledge about the underlying ground-truth weight tensor. This is achieved by defining a mode-$n$ graph which captures the corresponding intra-modal structure. For the proposed experiment, the mode-$n$ graph adjacency matrices, $\textbf{A}^{(n)} \in \mathbb{R}^{10 \times 10}$, $n=1,\ldots,4$, were computed such that
\begin{equation}
{a}^{(n)}_{i, j} = \texttt{exp} \{- \beta {\norm{\textbf{u}^{(n)}_{i} - \textbf{u}^{(n)}_{j}}}\}
\end{equation}
where $\textbf{u}^{(n)}_{i} \in \mathbb{R}^{5}$ and $\textbf{u}^{(n)}_{j} \in \mathbb{R}^{5}$ denote respectively the $i$-th and $j$-th row vector of the mode-$n$ factor matrix, ${\textbf{U}^{(n)}} \in \mathbb{R}^{10 \times 5}$. In this way, the entries of the mode-$n$ graph adjacency matrix encode the similarity between row vectors of the ground-truth mode-$n$ factor matrix. The mode-$n$ Laplacian matrices, $\textbf{L}^{(n)} \in \mathbb{R}^{10 \times 10}$, were then computed as discussed in Section \ref{sec:graph_sig}, which were used to regularize the learning process and to guide the estimated weights closer to the ground-truth.

\subsection{Models}

The proposed GRTR was evaluated against: (i) Linear Regression (LR), (ii) Linear Regression with L2 regularization (L2LR), (iii) standard Tensor Regression (TR), and (iv) Tensor Regression with L2 regularization (L2TR). For the LR and L2LR models, the ground-truth weight tensor was vectorized as $\textbf{w} = \vectorize \left( \mathcalbf{W} \right) \in \mathbb{R}^{10,000}$. 

\subsection{Results}

\begin{table}
\begin{center}

\begin{small}
\begin{tabular}{c c c c c c} 
\toprule

Model   & LR        & L2LR      & TR        & L2TR      & GRTR      \\ \midrule
MSE     & 0.137     & 0.137     & 0.083     & 0.051     & \textbf{0.025}    \\
TR-EVS  & \textbf{100.0\%}   & \textbf{100.0\%}   & 99.2\%    & 95.4\%    & 92.8\%   \\
TE-EVS  & -1.9\%    & -1.8\%    & 57.9\%    & 65.6\%    & \textbf{73.1\%}   \\
params  & 10,000    & 10,000    & \textbf{200}       & \textbf{200}       & \textbf{200}       \\
\bottomrule
\end{tabular}
\end{small}

\caption{Experiment results for the considered models. The first row shows the Mean Squared Error (MSE) bewteen the ground-truth weight tensor and the estimated weight tensor. The second and third rows show respectively the TRain set Explained Variance Scores (TR-EVS) and TEst set Explained Variance Scores (TE-EVS). The final row shows the number of parameters that need to be estimated as a measure of space complexity. The proposed GRTR model achieved the best results by a large margin while requiring only 200 parameters, while the other considered models over-fit the training set owing to the noisy and large-dimensional nature of the problem.}
\label{table:synthetic_res}
\end{center}
\end{table}

Simulation results are illustrated in Table \ref{table:synthetic_res}. Linear models, such as LR and L2LR, needed to estimate 10,000 parameters from 100 noisy samples, which led to over-fitting in the training set due to the noisy and large-dimensional nature of data. This resulted in negative explained variance scores in the test set and large mean-squared-errors in the estimation of the ground-truth weight tensor. Classical tensor models, such as TR and L2TR, were able to better capture the inherent multi-way relations in data by virtue of their tensor formulation, which reduced the space complexity to only 200 parameters. However, they still struggled due to large noise and small number of observations. The proposed GRTR, aided by the domain knowledge encoded in the mode-$n$ graphs, achieved the best overall scores by a large margin, while incurring the same small complexity cost of other tensor models.


\section{Experiments on Real Data} \label{sec:exp}

We next validate the proposed model in a real-world multi-way financial forecasting experiment.

\subsection{Data}

Stocks within the S\&P500 index between 2010 and 2020 were considered. Some stocks were removed due to data quality issues, leaving 450 stocks in total for the experiment. For each of these stocks, 6 daily market data features were obtained: (i) adjusted closing price, (ii) closing price, (iii) highest price, (iv) lowest price, (v) opening price, and (vi) trading volume. The raw market data were then processed to ensure stationary features by computing the log-returns, $r_t = log({p_t}) - log({p_{t-1}})$, where $p_t$ is the value of the market data at the $t$-th time-step. Using a rolling window of 5 time-steps, a set of tensor-structured inputs was generated at each point in time, $\textbf{X}^{(t)} \in \mathbb{R}^{T \times S \times F}$, where $F=6$ features for $S=450$ stocks were extracted over $T=5$ past time-steps. For each of the order-$3$ input tensors at a particular time-step, the corresponding output label was set to be the return of the S\&P500 index at the next time-step. The data was then partitioned using a standard $50\%$-$30\%$-$20\%$ train-validation-test split, and standardized to have zero mean and unit variance within the training set.

\subsection{{Domain Knowledge through Graphs}}

The performance of the proposed GRTR model will vary significantly depending on the design of the graph Laplacian matrix. Designing the optimal graph for equity returns forecasting is beyond the scope of the paper and is left to the domain experts. Instead, a sector graph was used in this experiment for illustrative purposes. 

Using the sector information from mode-$2$ (i.e., the stocks mode), a graph with $S$ vertices representing $S$ different stocks was constructed, where the edges connect a pair of stocks if they belong to the same sector. We then computed the corresponding Laplacian matrix for the mode-$2$, $\textbf{L}^{(2)} \in \mathbb{R}^{S \times S}$, as discussed in Section \ref{sec:graph_sig}. The proposed graph Laplacian matrix encodes knowledge about the sector information of the mode-$2$ stocks, which is expected to improve the learning process, as it guides the learning of the mode-$2$ factor matrix to be consistent with the correlation structure exhibited by the stock market.

\subsection{Models}

As in Section \ref{sec:exp_syn}, the proposed GRTR was evaluated against: (i) Linear Regression (LR), (ii) Linear Regression with L2 regularization (L2LR), (iii) standard Tensor Regression (TR), and (iv) Tensor Regression with L2 regularization (L2TR). For the LR and L2LR models, the ground-truth weight tensor was vectorized as $\textbf{w} = \vectorize \left( \mathcalbf{W} \right) \in \mathbb{R}^{13,500}$.

\subsection{Results}

The results for the proposed experiment are reported in Table \ref{table:reg_big}, which validates the proposed GRTR framework, as it obtained the best scores both in terms of computational complexity costs and out-of-sample accuracy scores. 

\begin{table}
\begin{center}

\begin{small}
\begin{tabular}{c c c c c c} 
\toprule

Model   & LR        & L2LR      & TR        & L2TR      & GRTR          \\ \midrule
TR-ACC  & \textbf{100.0\%}  & \textbf{100.0\%}  & 59.0\%    & 55.6\%    & 55.6\%        \\
TE-ACC  & 47.9\%    & 47.9\%    & 51.7\%    & 54.5\%    & \textbf{54.7\%}        \\
params  & 13,500    & 13,500    & \textbf{461}       & \textbf{461}       & \textbf{461}       \\
\bottomrule
\end{tabular}
\end{small}

\caption{Experimental results for the considered models. The first and second rows show respectively the TRain set ACCuracy (TR-ACC) and TEst set ACCuracy (TE-ACC) achieved by the considered models. The final row shows the required number of parameters (i.e., space complexity). The proposed GRTR model achieved the best results while requiring only 461 parameters, which is closely followed by the TR and the L2TR models. Linear models once again over-fitted the training set due to the noisy and large-dimensional nature of the problem.}
\label{table:reg_big}
\end{center}
\end{table}

\section{Interpretability} \label{sec:interpretability}

By virtue of multi-linear algebra, the GRTR model offers enhanced interpretability, both in terms of individual regression coefficients and different regression dimensions, as elaborated below.

\subsection{Coefficient-Wise Interpretability}

Similar to the classical linear regression paradigm, the proposed GRTR model allows us to recover the exact coefficient responsible for modelling a given input variable. For instance, the input value, $x_{i_1, i_2, \ldots, i_N}$ within $\mathcalbf{X} \in \mathbb{R}^{I_1 \times I_2 \times \cdots \times I_N}$, is directly linked to the coefficient, $w_{i_1, i_2, \ldots, i_N}$ of the weight tensor, $\mathcalbf{W} \in \mathbb{R}^{I_1 \times I_2 \times \cdots \times I_N}$, which can be reconstructed from its factor matrices according to (\ref{eq:cpd_contraction}).

\begin{example}
Consider the order-3 input tensor from the experiment in Section \ref{sec:exp}, $\mathcalbf{X} \in \mathbb{R}^{T \times S \times F}$. The effect of the input variable, $x_{1, 10, 2}$ (i.e., the $f=2$-nd feature of the $s=10$-th stock at the $t=1$-st past time-step), on the output, is directly linked to the corresponding entry in the weight tensor, $w_{1, 10, 2}$. In the proposed real data experiment, this value was equal to {\texttt{-0.00017}}.
\end{example}

\begin{figure*}
	\centering
	\includegraphics[width=1.0\linewidth]{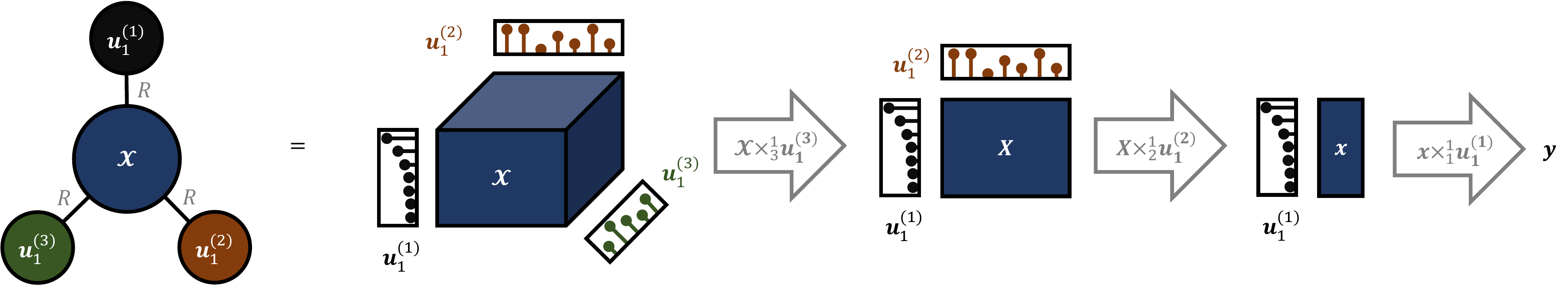}
	\caption{Illustration of a rank $R=1$ order $N=3$ tensor regression operation. The data tensor, $\mathcalbf{X} \in \mathbb{R}^{T \times S \times F}$, contracts with the factor vectors, $\textbf{u}_1^{(n)}$, $n=1,2,3$, to generate the overall scalar output, $y = \mathcalbf{X} \times_1^1 \textbf{u}_1^{(1)} \times_2^1 \textbf{u}_1^{(2)} \times_3^1 \textbf{u}_1^{(3)}$. Each mode-$n$ contraction represents a linear combination in the $n$-th subspace of $\mathcalbf{X} \in \mathbb{R}^{T \times S \times F}$, which reduces the order of the tensor until a scalar is obtained. In the above example, the original order-$3$ tensor is reduced to a matrix through a tensor-by-vector contraction (defined in Section \ref{sec:prelim}), which is then reduced to a vector through a matrix-by-vector multiplication, which is then reduced to a scalar through a vector-by-vector inner product.}
	\label{fig:dimensions_interpretability}
\end{figure*}

\subsection{Mode-Wise Interpretability}

By construction, the large weight tensor, $\mathcalbf{W}$, is decomposed by mode-wise factor matrices, $\textbf{U}^{(n)}$, which effectively breaks-down the large multi-dimensional regression problem into a set of smaller, physically meaningful, mode-wise sub-problems. This allows us to interpret the obtained model coefficients from a mode-wise point of view. 

To break down the tensor regression into a series of interpretable mode-wise operations, the tensor regression equation can be rewritten as a sum of tensor contractions by applying the rank-$1$ tensor inner product property discussed in Section \ref{sec:prelim_tensors}, as
\begin{equation} \label{eq:sum_of_contractions}
    \begin{aligned}
        y 
        &= \langle \mathcalbf{X}, \mathcalbf{W} \rangle + b \\
        &= \Big \langle \mathcalbf{X}, \sum_{r=1}^R \textbf{u}_r^{(1)} \circ \textbf{u}_r^{(2)} \circ \cdots \circ \textbf{u}_r^{(N)} \Big \rangle + b \\
        &= \Big \langle \mathcalbf{X}, \sum_{r=1}^R \bigcirc_{n=1}^N \textbf{u}_r^{(n)} \Big \rangle + b \\
        &= \sum_{r=1}^R \Big \langle \mathcalbf{X}, \bigcirc_{n=1}^N \textbf{u}_r^{(n)} \Big \rangle + b \\
        &= \sum_{r=1}^R \left( \mathcalbf{X} \times_1^1 \textbf{u}_r^{(1)} \times_2^1 \textbf{u}_r^{(2)} \times_3^1 \cdots \times_N^1 \textbf{u}_r^{(N)} \right) + b \\
    \end{aligned}
\end{equation}

Note that the computation of the tensor-by-vector contraction, $\mathcalbf{Z} = \mathcalbf{X} \times_n^1 \textbf{u}_r^{(n)}$, can be interpreted as performing a simple linear combination in the $n$-th subspace of $\mathcalbf{X} \in \mathbb{R}^{I_1 \times I_2 \times \cdots \times I_N}$, as discussed in Section \ref{sec:prelim_tensors}. Therefore, the sequence of contractions in equation (\ref{eq:sum_of_contractions}), $\mathcalbf{X} \times_1^1 \textbf{u}_r^{(1)} \times_2^1 \textbf{u}_r^{(2)} \times_3^1 \cdots \times_N^1 \textbf{u}_r^{(N)}$, can be interpreted as performing a sequence of linear combinations over every $n$-th subspace of $\mathcalbf{X} \in \mathbb{R}^{I_1 \times I_2 \times \cdots \times I_N}$. 



\begin{figure} 
	\centering
	\includegraphics[width=1.0\linewidth]{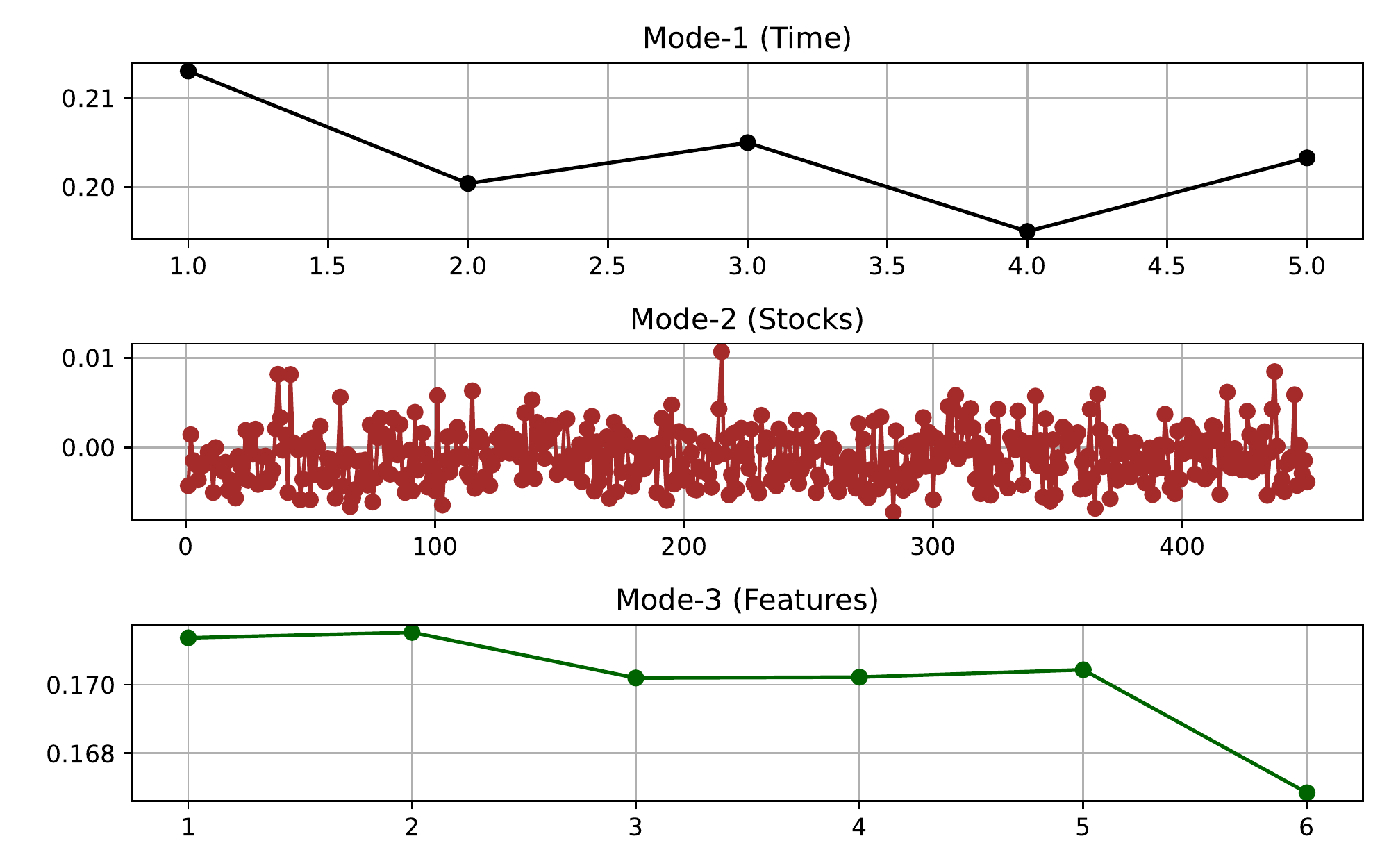}
	\caption{Estimated rank $R=1$ tensor-regression factor vectors from the proposed experiment. A factor vector is estimated for the time mode, $\textbf{u}^{(1)} \in \mathbb{R}^T$, the stock mode, $\textbf{u}^{(2)} \in \mathbb{R}^S$, and feature mode,  $\textbf{u}^{(3)} \in \mathbb{R}^F$. Note that the factor vector values for the time-mode can be interpreted as a weighted moving average operation in the time-subspace, as larger weighting are placed towards more recent data ($t=1$) compared to older data ($t=5$).}
	\label{fig:interpretability_dims}
\end{figure}

\begin{example} 
Consider the order-3 data tensor from the experiment in Section \ref{sec:exp}, $\mathcalbf{X} \in \mathbb{R}^{T \times S \times F}$. For $R=1$, equation (\ref{eq:sum_of_contractions}) simplifies to one single contraction, $y = \left( \mathcalbf{X} \times_1^1 \textbf{u}_1^{(1)} \times_2^1 \textbf{u}_1^{(2)} \times_3^1 \textbf{u}_1^{(3)} \right) + b$, which can be visualized as in Figure \ref{fig:dimensions_interpretability}. Therefore, tensor regression performs three linear combinations, one in each mode. The time weights, $\textbf{u}_r^{(1)} \in \mathbb{R}^T$, the stocks weights, $\textbf{u}_r^{(2)} \in \mathbb{R}^S$, and the feature weights, $\textbf{u}_r^{(3)} \in \mathbb{R}^F$, determine respectively the linear weighted contribution of all $T$ time-steps, $S$ stocks, and $F$ features to the final regression output. For illustration, Figure \ref{fig:interpretability_dims} shows the estimated weights of the GRTR model with rank $R=1$. In particular, the time-domain weights can be interpreted as a weighted moving average operation over the time-dimension, with more weighting placed towards recent data (time-step $t=1$) compared to older data (time-step $t=5$). 



\end{example}

\section{Conclusion} \label{sec:conc}
We have introduced a tensor regression framework with graph-based regularization, which makes it possible to efficiently incorporate domain knowledge into large-scale, multi-dimensional regression through Laplacian graph regularization. The so introduced model has been shown to be physically meaningful and to provide improved results against standard vector and tensor regression models, both in terms of out-of-sample performance and computational costs. By virtue of tensor algebra, the proposed model is also shown to enable enhanced interpretability and control over the learning process. The proposed framework is general and opens avenues for exploring numerous possibilities for formulating economically meaningful graphs from cross-asset variables; these can also can guide learning models towards higher performance and meaningful predictions. 

\bibliographystyle{IEEEbib.bst}
\bibliography{references.bib}

\end{document}